\documentstyle[aps,preprint]{revtex}
\begin{document}
\draft
\title{Statistical mechanics of macromolecular networks
without replicas}
\author{Michael P Solf, Thomas A Vilgis}
\address{Max-Planck-Institut f\"ur Polymerforschung,
Postfach 3148, 55021 Mainz, Germany} 
\date{to appear in J. Phys. A: Math. Gen. {\bf 28}, December (1995)}
\maketitle

\begin{abstract}
We report on a novel approach to the Deam-Edwards model for
interacting polymeric networks without using replicas. Our
approach utilizes the fact that a network modelled from a
single non-interacting Gaussian chain of macroscopic size
can be solved exactly, even for randomly distributed
crosslinking junctions.  We derive an {\it exact}
expression for the partition function of such a generalized
Gaussian structure in the presence of random external
fields and for its scattering function $S_0$.  We show that
$S_0$ of a randomly crosslinked Gaussian network (RCGN) is
a self-averaging quantity and depends only on crosslink
concentration $M/N$, where $M$ and $N$ are the total
numbers of crosslinks and monomers.  From our  derivation
we find that the radius of gyration $R_{\mbox{\tiny g}}$ of
a RCGN is of the universal form $R_{\mbox{\tiny g}}^2=(0.26
\pm 0.01)a^2 N/M$, with $a$ being the Kuhn length.  To
treat the excluded volume effect in a systematic,
perturbative manner, we expand the Deam-Edwards partition
function in terms of density fluctuations analogous to the
theory of linear polymers.  For a highly crosslinked
interacting network we derive an expression for the free
energy of the system in terms of $S_0$ which has the same
role in our model as the Debye function for linear
polymers.  Our ideas are easily generalized to crosslinked
polymer blends which are treated within a modified version
of Leibler's mean field theory for block copolymers.
\end{abstract}

\pacs{61.41.+e, 61.43.-j, 82.20.Db}

\section{Introduction}

Randomly crosslinked macromolecules present a challenging
field from a physical as well as from a mathematical point
of view with many practical applications in polymer
sciences. For this it is very unfortunate that the
statistical mechanics of polymer networks is still poorly
understood.  Although there has been a great deal of
theoretical interest in this topic over the last couple of
years
\cite{deamed,edwa,badoed,golgol,golzip,ipnvil,vilsol,viled},
we feel that a satisfying answer of how to deal with
polymer networks from a microscopic point of view is still
missing. One of the reasons is that most of the recent work
on the subject \cite{golgol,golzip} is based on replica
field theory originally introduced into polymer science by
Edwards {\it et al} \cite{deamed,edwa,badoed} in which,
however, much of the underlying physics remains hidden in
the complicated mathematics of the replica formalism
\cite{spingl}.

It is the purpose of this paper to present an alternative
approach to the statistical mechanics of randomly
crosslinked macromolecules that goes beyond the earlier
phantom type models (for details see, for example,
reference \cite{trelo}), but at the same time avoids the
well known difficulties associated with the replica trick
\cite{spingl}.  Our theory is based on the minimal network
model by Deam and Edwards
\cite{deamed,edwa,badoed,golgol,golzip,ipnvil,vilsol},
however, avoids replica field theory completely. The
strategy in this paper is outlined below: As
generalization of the Wiener measure in the theory of
linear polymers \cite{doied} we introduce the concept of a
randomly crosslinked Gaussian network (RCGN) in section 2.
As we will show in section 3, the non-interacting problem
can be solved exactly, even for random crosslinking
junctions. In section 3 we develop our general formalism
and derive an exact formula for the partition function of a
generalized Gaussian structure in the presence of random
external fields.  This equation (20) is the central
mathematical result of the paper, for it is also a
generating function from which further results are
obtained.  In section 4 we consider some applications. In
particular we calculate the static structure function $S_0$
for RCGNs without excluded volume interaction.  The
important finding here is that $S_0$ is a self-averaging
quantity, i.e., it does not depend on the topological
details of the model. The consequence is that for RCGNs
$S_0$ is a quasi universal function that has a similar role
for polymer networks as the Debye function for linear
polymers. The interacting case is treated in section 5.  To
take into account excluded volume interaction, we transform
our original network Hamiltonian to collective density
variables. We work here in close analogy with the excluded
volume problem for dense polymer melts \cite{doied}. For
highly crosslinked systems it is sufficient to consider
only lowest order density fluctuations, although higher
order terms are readily calculated within our formalism. In
this case it is easy to show that the free energy $F$ can
be expressed in terms of the scattering function $S_0$ of
the non-interacting system and the excluded volume
parameter. It is crucial to realize that $F$ depends on the
frozen degrees of freedom (random crosslinking junctions)
only via $S_0$, and no further quenched averaging remains
to be done. To treat crosslinked polymer blends we apply a
modified version of Leibler's theory for microphase
separation in block copolymers \cite{leibler} and show how
to obtain similar criteria for phase instability in
multi-component networks.  Generalizations and outlook are
given in section 6.  However, more detailed calculations of
the latter applications will be presented in a forthcoming
publication.

\section{Formulation of the model}

We consider flexible interacting macromolecules on the
level of the Edwards Hamiltonian \cite{deamed}. For a
single polymer chain in $d$ spatial dimensions the Edwards
Hamiltonian consists of two parts $H=H_{\mbox{\tiny
W}}+H_{\mbox{\tiny I}}$, where
\begin{equation}
\beta H_{\mbox{\tiny W}}=\frac d{2a^2}\sum_{i=1}^N({\bf
R}_i-{\bf R}_{i-1})^2
\end{equation}
is the Wiener measure that models the connectedness of the
chain as a Gaussian. Self-avoidance between monomers is
described by a pseudopotential of the form
\begin{equation}
\beta H_{\mbox{\tiny I}}=v\sum_{0\leq i<j}^N\delta \left(
{\bf R}_i-{\bf R}_j\right) ~, 
\end{equation}
where $\delta ({\bf R})$ is the Dirac delta function. In
(1) and (2) we have adopted the following notation: $v>0$
is the second virial coefficient \cite{doied} that
characterizes the repulsive excluded volume interaction
between monomers, $N$ is the degree of polymerization, $a$
the Kuhn length, and ${\bf R}_i$ ($i=0,..,N$) are monomer
coordinates; $\beta =(k_{\mbox{\tiny B}}T)^{-1}$ as usual.
For convenience we restrict the following discussion to
networks that are modelled  from one single but huge
polymer chain of macroscopic dimension (figure 1). It was
shown that this simplification gives the correct physics
for highly crosslinked polymer networks above the
percolation threshold \cite{deamed}.  A generalization of
our method to multi-polymer networks will be discussed
later on in section 6.

To describe $M$ permanently crosslinked monomers, we
specify  each junction by a pair of randomly chosen
``crosslink coordinates'' $i_e,j_e$ ($0\leq i_e,j_e \leq
N$, $e=1,...,M$), such that monomer ${\bf R}_{i_e}$ is
connected to monomer ${\bf R}_{j_e}$ (figure 1). The whole
set of junction points $\mbox{C}=\{i_e,j_e\}_{e=1}^M$
represents the random connectivity of the network. Within
the framework of the Deam-Edwards model \cite{deamed} the
partition function of a Gaussian phantom network with
excluded volume is given by
\begin{equation}
Z(\mbox{C})=\int\limits_V\prod_{i=0}^Nd{\bf R}_i
\,e^{-\beta (H_{\mbox{\tiny W}}+H_{\mbox{\tiny I}})}\,
\prod_{e=1}^M\delta \left( {\bf R}_{i_e}-{\bf
R}_{j_e}\right )~, 
\end{equation}
where total phase space is now restricted by the additional
crosslinking constraints. Equation (3) describes a phantom
network in a sense that the polymer chain is free to pass
through itself irrespective of entanglements. Chain motion
is only restricted by the presence of permanent crosslinks
and the excluded volume interaction.

In the replica formalism the next step is to perform a
quenched average over the logarithm of $Z(\mbox{C})$ with a
suitable distribution for the ``frozen'' crosslink
coordinates $\{i_e,j_e\}_{e=1}^M$. Using the replica trick
\cite{spingl} this leads to a non-trivial modification of
the interaction term $H_{\mbox{\tiny I}}$ in which all
replicas become coupled
\cite{deamed,edwa,badoed,golgol,golzip}.  Applying standard
techniques for setting up field theories in polymer
physics, we were able to map equation (3) into a $dn$
dimensional O$(m)$ field theory in the limit
$n,m\rightarrow 0$ \cite{vilsol}. Unfortunately in the
replica formalism further progress highly relies upon crude
approximations or variational assumptions
\cite{deamed,edwa,badoed,golgol,golzip,ipnvil,vilsol}.\\

Contrary to the replica method we do not carry out the
quenched average at this stage of the calculation. Instead
we model the delta function in (3) by a Gaussian
distribution with width $\varepsilon $ in the limit
$\varepsilon \rightarrow 0$. Therefore we are keeping all
random crosslink coordinates explicitly in the partition
function. Formally this means that if the Wiener measure
$\beta H_{\mbox{\tiny W}}$ is replaced by the more general
expression for a RCGN
\begin{equation}
\beta H_{\mbox{\tiny G}}=\frac d{2a^2}\sum_{i=1}^N({\bf
R}_i-{\bf R}_{i-1})^2+\frac {d}{2\varepsilon
^2}\sum_{e=1}^M({\bf R}_{i_e}-{\bf R}_{j_e})^2~, 
\end{equation}
we can eliminate the delta constraint in (3). With
equations (2) and (4) we are now in a position to introduce
our RCGN Hamiltonian as follows
\begin{equation}
H=H_{\mbox{\tiny G}}+H_{\mbox{\tiny I}}~.
\end{equation}
To enforce the crosslinking constraints we finally have to
perform the limit $\varepsilon \rightarrow 0$. It is
mathematically convenient to normalize the Gaussian measure
defined by (4) with respect to the non-interacting system
which will serve us as a reference state. In this case the
partition function in (3) can be cast into the more
convenient form
\begin{equation}
Z(\mbox{C})=\left\langle e^{-\beta H_{\mbox{\tiny
I}}}\right\rangle _0= \exp \Big (-\beta (F-F_0) \Big )~,
\end{equation}
where the average $\left\langle ...\right\rangle _0$ stands
for
\begin{equation}
\lim _{\varepsilon \rightarrow 0}\Big (
\int \prod_{i=0}^Nd{\bf R}_i\,e^{-\beta H_{\mbox{\tiny
G}}}...\Big ) \times
\Big( \int \prod_{i=0}^Nd{\bf R}_i\,e^{-\beta
H_{\mbox{\tiny G}}}\Big )^{-1}~.
\end{equation}
For $v=0$ the free energy $F$ of the interacting network
reduces to that of a non-interacting RCGN denoted by $F_0$.
The partition function (6) is completely equivalent to the
one used by Edwards {\it et al} \cite{deamed,edwa,badoed}.
It is also identical to the one used in more recent works
by Goldbart and coworkers \cite{golgol,golzip}, except for
the fact that in this paper we are dealing with a
single-chain network exclusively. This is primarily a
matter of convenience which has been discussed in greater
detail in references \cite{deamed,edwa,badoed}. The
Hamiltonian (5) is of course easily generalized to
multi-chain or multi-component networks.

To make further analytic progress, it will become useful to
decouple the interaction term $H_{\mbox{\tiny I}}$ by
rewriting (2) in terms of collective density variables
$\Phi _{{\bf k}}=\sum_{i=0}^N\exp( i{\bf k}{\bf R}_i)$ and
applying the Hubbard-Stratonovich transformation to (6).
Since this is standard procedure \cite{golgol,doied} we
quote here only the final result
\begin{eqnarray}
Z(\mbox{C}) & \propto \int \prod_{{\bf k}>0} d\Phi_{{\bf
k}}\, \exp\Big(-\frac{v}{V}\sum_{{\bf k}>0} \Phi_{{\bf k}}
\Phi_{-{\bf k}} \Big)\\
& \times \Big\langle \exp \Big( i
\frac{v}{V}\sum_{i=0}^N \sum_{{\bf k}\neq 0} \Phi_{-{\bf
k}} \exp (i {\bf k} {\bf R}_i) \Big)\Big\rangle_0
~.\nonumber 
\end{eqnarray}
By ${\bf k}>0$ we mean the positive half of ${\bf k}$
space, i.e., $k_x>0$, and $V$ is the volume. The partition
function (8) in combination with the measure defined by (7)
is equivalent to the  minimal network model of Deam and
Edwards \cite{deamed} and will be our starting point for
further investigations.

\section{General formalism}

Before dealing with the more complicated excluded volume
situation directly, it is easier to develop the
mathematical formalism for non-interacting RCGNs at first.
To evaluate equation (8) in a systematic fashion, it is
standard  to expand the first exponential in the entropic,
second part of (8) in terms of density fluctuations
$\Phi_{-{\bf k}}$ or the excluded volume parameter. In any
case the main mathematical task is to calculate averages
which are of the general form $\big\langle \exp (i {\bf
b}^t {\bf R})\big\rangle_0$, where we have introduced
$d\times (N+1)$ dimensional ``supervectors'' ${\bf R}=({\bf
R}_0,{\bf R}_1,...,{\bf R}_N)^t$ and ${\bf b}=({\bf
b}_0,{\bf b}_1,...,{\bf b}_N)^t$. For the moment we assume
that ${\bf b}$ is completely arbitrary, but does not depend
on ${\bf R}$ explicitly. By ${\bf b}_i {\bf R}_j$ we will
always mean a $d$ dimensional inner vector product,  and
$t$ denotes the transposed vector.  Since $\left\langle
...\right\rangle_0$ involves a Gaussian integration with
the measure defined by (7), it is essential to find the
inverse matrix of the quadratic form, equation (4), first.
For this it is most convenient to switch to  matrix
notation.

Using  matrix notation in which ${\bf R}$ is the above
defined ``supervector'', the quadratic form in (4) can be
written as follows
\begin{equation}
\beta H_{\mbox{\tiny G}}=\frac{d}{2\varepsilon^2}~{\bf R}^t
\Big( z{\cal W}+\sum_{e=1}^M {\cal K}(i_e,j_e) \Big){\bf R}~,
\end{equation}
where $z=(\varepsilon /a)^2$, and symmetric $(N+1) \times
(N+1)$ matrices ${\cal W}$ and ${\cal K}$. Here
\begin{equation}
{\cal W}=\left(
\begin{array}{rrrrr}
1 & -1 & 0 & \cdots & 0 \\ -1 & 2 & -1 & \cdots & 0 \\
\vdots & \ddots & \ddots & \ddots & \vdots \\ 
0 & \cdots & -1 & 2 & -1 \\ 0 & \cdots & 0 & -1 & 1
\end{array}
\right ) 
\end{equation}
denotes the connectivity matrix associated with
$H_{\mbox{\tiny W}}$, and
\begin{equation}
{\cal K}(i_e,j_e)=\left(
\begin{array}{rrcrr}
0 & 0 & 0 & 0 & 0 \\ 0 & 1 & \vdots & -1 & 0 \\ 0 & \cdots
& 0 & \cdots & 0 \\ 0 & -1 & \vdots & 1 & 0 \\ 0 & 0 & 0 &
0 & 0 
\end{array}
\right ) 
\begin{array}{c}
\vdots \\ 
\leftarrow i_e
\mbox{-th position} \\ \vdots \\ 
\leftarrow j_e
\mbox{-th position} \\ \vdots 
\end{array}
\end{equation}
represents a single crosslinking junction at random
position $(i_e,j_e)$.  Equation (9) is easily verified by
inspection.

For the following it is essential to note that the
quadratic form, equation (9), is only {\it semi}positive
definite, hence no matrix inverse does exist. The
non-negativity of $H_{\mbox{\tiny G}}$ follows immediately
from equation (4), since it is a sum of squares.  However,
there is one zero eigenvalue associated with eigenvector
${\bf R}=({\bf 1},...,{\bf 1})^t$. This is easily seen from
noting that the column sum of the matrix in (9) is always
zero irrespective of the value of $z$. Before we proceed,
we need to transform (9) to a positive definite quadratic
form by removing the mode which belongs to eigenvalue zero.
This mode corresponds to a displacement of the center of
mass, and since we are dealing with a simply-connected
structure, there can be only {\it one} such mode. The
desired transformation is easily accomplished by switching
to internal coordinates ${\bf r}_i={\bf R}_i-{\bf
R}_{i-1}$, or in matrix notation ${\bf R}={\cal D}\tilde
{{\bf r}}$, where
\begin{equation}
\tilde {{\bf r}}=\left(
\begin{array}{c}
{\bf R}_0 \\ {\bf r}_1\\ \vdots \\ {\bf r}_N
\end{array}
\right)
~~~\mbox{and}~~~ {\cal D}=\left(
\begin{array}{cccc}
1 & 0 & \cdots  & 0 \\ 1 & 1 & \cdots  & 0 \\
\vdots  & \vdots  & \ddots  & \vdots  \\ 
1 & 1 & \cdots  & 1
\end{array}
\right) ~.
\end{equation}
For the multi-chain network we refer to our discussion in
section 6. With the above manipulations the measure
corresponding to equation (9) takes on the simple form
\begin{equation}
Z_0(\mbox{C})=\int \prod_{i=0}^Nd{\bf R}_i\,e^{-\beta
H_{\mbox{\tiny G}}} 
= V\int \prod_{i=1}^Nd{\bf r}_i\,\exp \Big( -\frac
d{2\varepsilon 
^2}\,{\bf r}^t{\cal M}\,{\bf r}\Big )~,
\end{equation}
where ${\bf r}=({\bf r}_1,...,{\bf r}_N)^t$, and we
performed an integration over ${\bf R}_0$.  In the limit
$\varepsilon \rightarrow 0$ equation (13) yields the
partition function of a RCGN without excluded volume. The
$N$ dimensional matrix ${\cal M}$ in (13) is given by
\begin{equation}
{\cal M}(z)=z{\cal I}+{\cal P}{\cal P}^t~,
\end{equation}
where ${\cal I}$ denotes an $N$ dimensional unit matrix.
The outer matrix product $ {\cal P}{\cal P}^t$ is formed
with the $N\times M$ ``crosslink matrix'' ${\cal P}=({\bf
p}_1,...,{\bf p}_M)$, where each  column vector is defined
by
\begin{equation}
{\bf p}_e=(0,\dots ,0,\underbrace{1,1,\dots ,1,1}_{\mbox{
$i_e+1$ to $j_e$}},0,\dots ,0)^t~~,~e=1,...,M~.
\end{equation}
The 1's in (15) run from the $(i_e+1)$th to the $j_e$th
position, the rest of the components is 0. For convenience
we will assume that $j_e$ is always larger than $i_e$.
Equation (14) is derived in appendix A.

The crosslink matrix ${\cal P}$ is defined in such a way
that it contains complete information about the crosslink
positions in an unique way.  By construction it is exactly
this extra term in ${\cal M}$ that distinguishes the
network problem from a linear polymer. For the following
derivation it is crucial to have the crosslinking
constraint in equation (14) in form of an {\it outer}
matrix product of ${\cal P}$.  Note that ${\cal M}$ is
positive definite, and thus its inverse exists.

Using equations (13) and (14) the measure in (7) can be
redefined in terms of ${\cal M}$
\begin{equation}
\big\langle ...\big\rangle_0 = \lim_{\varepsilon\rightarrow
0} \frac{1}{Z_0(\mbox{C})}\int \!{\bf R}_0 \!\int
\!\prod_{i=1}^N d{\bf r}_i \, \exp \Big(
-\frac{d}{2\varepsilon^2}\, {\bf r}^t{\cal M}\,{\bf r}\Big 
)\,...
\end{equation}
In the remainder of this section we will show that this
average exists for $\varepsilon \rightarrow 0$ and derive
an exact expression.

With (16) we are in a position to calculate expectation
values of the form $\big\langle \exp (i {\bf b}^t {\bf
R})\big\rangle_0$. Since ${\bf r}$ depends only linearly on
${\bf R}$, it is sufficient to consider expressions of the
form
\begin{equation}
\Big\langle \exp (i{\bf c}^t{\bf
r})\Big\rangle_0=\lim_{\varepsilon 
\rightarrow 0}\,\exp \Big( -\frac{\varepsilon ^2}{2d}\,{\bf
c}^t{\cal M} ^{-1}(z)\,{\bf c}\Big)~,
\end{equation}
where ${\bf c}$ is again some arbitrary vector which does
not explicitly depend on ${\bf R}$. We can always go back
to the original monomer coordinates since $\tilde{\bf
r}={\cal D}^{-1}{\bf R}$. The key problem here is to find
the inverse of ${\cal M}(z)$ in the limit $z=(\varepsilon
/a)^2 \rightarrow 0$ which depends on all the crosslink
coordinates $\{i_e,j_e\}_{e=1}^M$. This is accomplished by
invoking an identity due to Frobenius, Schur and Woodbury
\cite{zurfa,lewnew}. An alternative but shorter proof of
the theorem is given in appendix B. The important finding
is that
\begin{equation}
{\cal M}^{-1}(z)=\frac 1z\Big( {\cal I}-{\cal P}({\cal
P}^t{\cal P})^{-1}{\cal P}^t\Big) ~.
\end{equation}
Making use of the fact that ${\cal P}^{+}=({\cal P}^t{\cal
P})^{-1}{\cal P}^t$ is a pseudoinverse of ${\cal P}$, i.e.
${\cal P}^{+}{\cal P}= {\cal I}$, it is trivial  to show
that ${\cal P}{\cal P}^{+}$ is a projector for ${\cal P}$.
It is also possible to prove the following alternative
representations \cite{raomit}
\begin{equation}
{\cal P}{\cal P}^{+}={\cal X}{\cal X}^t=\sum_{e=1}^M{\bf
x}_e^{}{\bf x}_e^t ~.
\end{equation}
Here ${\cal X}=({\bf x}_1,...,{\bf x}_M)$ is any
orthonormal vector basis ${\bf x}_e$, $e=1,...,M$ for the
$M$ dimensional vector space spanned by the ${\bf p}_e$'s
in (15). Combining equations (17--19) our central result of
this section can now be summarized by the following
formulas
\begin{eqnarray}
\Big\langle \exp (i{\bf c}^t{\bf r})\Big\rangle_0&=&\exp
\Big(- \frac{a^2}{2d} \,{\bf c}^t({\cal I}-{\cal X}{\cal
X}^t)\,{\bf c}\Big) \\ 
&=&\exp \bigg ( -\frac{a^2}{2d} \,\Big({\bf
c}^2-\sum_{e=1}^M ({\bf c}^t{\bf x}_e)^2 \Big ) \bigg
)~.\nonumber 
\end{eqnarray}
It is remarkable that equation (20) is of very simple and
special form although the crosslink coordinates
$\mbox{C}=\{i_e,j_e\}_{e=1}^M$ are completely random.
${\cal I}-{\cal X}{\cal X}^t$  is orthogonal to ${\cal P}$
and idempotent which can be seen by inspection. As a
consequence the only eigenvalues of ${\cal I}-{\cal X}{\cal
X}^t$ are 0 and 1 with degeneracies $M$ and $N-M$, and the
quadratic form in (20) is semipositive definite.  This
assures that the exponent in (20) is never positive.  The
non-triviality of the network problem enters the
calculation in form of ${\cal X}$ which can be found either
by orthonormalizing  the ${\bf p}_e$'s in (15) or directly
from (18).  The former is usually accomplished by
Gram-Schmidt orthonormalization \cite{zurfa} or numerically
by singular value decomposition \cite{numrec}. Note that
equation (18) only requires inversion of an $M\times M$
matrix, whereas ${\cal M}^{-1}$ is $N$ dimensional with
$N\gg M$ for a real network.

It is worthwhile to mention that equation (20) is also the
partition function of a Gaussian structure (random or not
depending on the choice of crosslinks
$\{i_e,j_e\}_{e=1}^M$) in the presence of random external
fields ${\bf b}_1,...,{\bf b}_N$. This is easily seen by
making the transformation in (20) ${\bf c}={\cal D}^t {\bf
b}$ back to the original monomer coordinates ${\bf R}_i$.
In the following section we consider more applications of
equation (20) with emphasis on RCGNs.

\section{Non-interacting random networks}

The reference quantity in the theory of linear polymers and
polymer melts \cite{doied,leibler} is the static structure
function  of a single non-interacting polymer chain. For a
RCGN it is given by
\begin{equation}
S_0({\bf k},\mbox{C})=\Big\langle
\sum_{i,j=0}^N \exp \Big(i{\bf k}({\bf R}_i -{\bf R}_j)\Big
)\Big\rangle_0~.
\end{equation}
The structure function  can be measured directly in polymer
solutions under $\Theta$-conditions via neutron scattering.
Physically it is the Fourier transform of the pair
correlation function.

From equation (20) it is  easy to derive an exact
expression for $S_0({\bf k},\mbox{C})$ by setting
\begin{equation}
{\bf c}={\bf k}\,
(0,\dots,0,\underbrace{1,1,\dots,1,1}_{\mbox{ $i+1$ to 
$j$}},0,\dots,0)^t~.
\end{equation}
From equations (20--22) we find
\begin{equation}
S_0({\bf k},\mbox{C})=\sum_{i,j=0}^N \exp \Big(
-\frac{k^2a^2}{2d} \big(|i-j|-({\bf y}_i-{\bf y}_j)^2
\big)\Big ). 
\end{equation}
The $M$ dimensional vectors ${\bf y}_i$ ($i=0,...,N$) are
given in terms of the orthonormal basis ${\cal X}$ in (19)
\begin{equation}
({\bf y}_1,...,{\bf y}_{N})=({\cal D}{\cal X})^t ~,
\end{equation}
where ${\cal D}$ is the lower triangular matrix defined in
(12), and ${\bf y}_0={\bf 0}$. Each ${\bf y}_i$ is a vector
whose components depend on the whole set of crosslink
coordinates C via ${\cal X}$. Equation (23) is of similar
structure as the result in reference \cite{warner} for a
RCGN under external stress obtained by completely different
means. Derivations of structure factors that are based on
the affine deformation hypothesis \cite{trelo} can be found
in references \cite{warner,pearson} but will not be dealt
with in this work.

Although our primary interest here are RCGNs, we can apply
equation (23) to problems with non-random connectivities as
well. As a simple example which can be solved analytically
and to illustrate our formalism, we consider a flexible
ring polymer as a trivial example of a non-random network
with only one crosslink. For a closed loop the crosslink
connects monomer ${\bf R}_0$ with ${\bf R}_N$, and the
crosslink matrix ${\cal P}$ is of the simple form ${\cal
P}={\bf p}_1=(1,1,\dots,1)^t$.  Thus ${\cal X}={\bf
x}_1=1/\sqrt{N}\,(1,1,\dots,1)^t$ and with the definition
in (24), ${\bf y}_i=i/\sqrt{N}$ for $i=0,...,N$. From
equation (23) we get the exact result
\begin{equation}
S_0({\bf k},\mbox{Ring})=\sum_{i,j=0}^N \exp \bigg (
-\frac{k^2a^2|i-j|}{2d} \Big (1-\frac{|i-j|}{N} \Big)\bigg
)~. 
\end{equation}
There are more cases in which ${\cal X}$ can be obtained
analytically.  Among these are star or branched polymers
and networks or manifolds with regular, non-random
connectivity. More details of these applications will be
given elsewhere. Since equation (23) is exact, it also
reduces for $M=0$ to the well known Debye function for
linear polymers
\cite{doied}.\\

We now specify our discussion to networks with quenched
random connectivity. For an arbitrary set of crosslinks we
are not able to derive analytic expressions for ${\bf y}_i$
in (23).  Moreover, within the framework of replica theory
macroscopic physical observables like $S_0$ are to be
averaged over the ``frozen'' variables, here all possible
sets of crosslink coordinates C.  To make further progress
it is therefore absolutely crucial to understand that in
the thermodynamic limit when $N$ and $M$ are sufficiently
large {\it any} specific but random crosslink configuration
C would produce the same result for $S_0$ if, for example,
equation (23) could be evaluated analytically. In the
literature this is well known as self-averaging.  In fact
for any self-averaging quantity like the free energy,
structure factor or radius of gyration (these are the
quantities we are mainly interested in) performing the
quenched average at the end of the calculation is not an
absolute necessity. To obtain $S_0$ for RCGNs it is
therefore sufficient to pick one specific set of random
numbers $0\leq i_e,j_e\leq N$, $e=1,...,M$, from a suitable
distribution function and evaluate (23) for this specific
but random connectivity. With the results from section 3 it
is indeed very easy to obtain ${\bf y}_i$ and carry out the
summation in (23) numerically (figure 2). To create a
specific set of crosslink coordinates
$\mbox{C}=\{i_e,j_e\}_{e=1}^M$ we have chosen an uniform
distribution $P(\mbox{C})=(1/N)^{2M}$ for simplicity
\cite{golgol}.  Physically this means that the frozen
crosslinks can in principle be anywhere along the chain
without restrictions.

Self-averaging is demonstrated in figure 2 for two random
networks with different crosslink configurations C and also
different network sizes $N$ and $M$. We find that
self-averaging is almost perfectly fulfilled even for
relatively small networks with $N \stackrel{\textstyle
>}{\sim} 5000$ and crosslink densities of a few percent.
Small deviations between the two networks in figure 2 are
due to finite size effects. The structure function depicted
in figure 2 is a universal function in a sense that any
other RCGN with $N \stackrel{\textstyle >}{\sim} 5000$ led
to identical curves.  As one would have expected $S_0$ does
only depend on the number of crosslinks $M$ and monomers
$N$  and {\it not} on microscopic details of crosslink
positions. In fact our exact calculation shows that for
RCGNs, $S_0({\bf k},\mbox{C}=\{i_e,j_e\}_{e=1}^M)=S_0({\bf
k},c)$ and that it depends only on crosslink concentration
$c=M/N$. The important consequence is that for RCGNs the
structure function $S_0({\bf k},c)$ can be viewed as the
equivalent to the Debye function in the theory of linear
polymers. By construction it is this complete analogy
between the theory of linear polymers and our model that
will allow us to approach the excluded volume problem in
the next section very similar as for linear polymers.

As a by-product of equation (23) we obtain an exact
expression for the radius of gyration $R_{\mbox{\tiny g}}$
of a RCGN. It is easy to show
\cite{doied} that in the limit ${\bf k}\rightarrow {\bf 0}$,
$S_0({\bf k},c)/N^2=1-{\bf k}^2 R_{\mbox{\tiny g}}^2/d$,
and from our numerical calculation we find that
$R_{\mbox{\tiny g}}^2=0.26a^2/c$, where $c=M/N$. The latter
result is also universal in a sense that the prefactor
$0.26\pm 0.01\approx 1/f$ was found for all networks with
functionality $f=4$ which were modelled from a uniform
crosslink distribution. Small fluctuations were mainly due
to the finite size of the networks under investigation.
Similar results have been suggested throughout the
literature. However, these calculations had to rely upon
various approximation schemes. For very large ${\bf k}$
values of the order $1/a$, $S_0({\bf k},c)$ decays as
$1/{\bf k}^2$ as expected for a non-interacting system.  In
the intermediate ${\bf k}$ range no simple power-law decay
could be found.  From the semipositive definiteness of (20)
and (23) it follows that the scattering function of any
simply-connected Gaussian structure is a monotonously
decreasing function in ${\bf k}^2$ with a maximum at ${\bf
k}={\bf 0}$. So far further analytic progress in (23)
depends on simplifying assumptions or approximations about
the form of ${\bf y}_i$'s in (24).

As another illustration of (20) we consider a generalized
Gaussian structure in a homogeneous electric field ${\bf
E}$ when each monomer is randomly charged with $q_i=\pm q$.
In this case the electric energy is given by $-\sum_i q_i
{\bf E}{\bf R}_i$, and we can directly get the partition
function from (20). Performing the quenched average over
$q_i$ which is analytically possible the free energy
becomes $F({\bf E})=F({\bf 0})-(a^2q^2{\bf
E}^2/2dk_{\mbox{\tiny B}}T)\,\mbox{Tr}\,{\cal Y}$, where
$\mbox{Tr}$ means trace, and ${\cal Y}={\cal D}({\cal
I}-{\cal X}{\cal X}^t) {\cal D}^t$. Applying the definition
in (24) we find that $\mbox{Tr}\,{\cal
Y}=N(N-1)/2-\sum_{i=1}^N {\bf y}_i^2$ and from a numerical
calculation $\mbox{Tr}\,{\cal Y}\approx 0.5 N/c$.

\section{Random networks with excluded volume}

\subsection{Free energy}

We want to outline how the free energy of a RCGN with
excluded volume interaction can be obtained without
resorting to replica methods. For the above considered
network the partition function (8) is now expanded in terms
of density fluctuations $\Phi_{-{\bf k}}$ up to second
order. This approximation is known to be valid in dense
systems when fluctuations about the mean density are small
\cite{doied}. In the Gaussian approximation \cite{doied} it
is straightforward to obtain the free energy density
$F(v,c)$ of the interacting system in terms of $S_0({\bf
k},c)$ of the non-interacting system
\begin{equation} 
F(v,c)-F(0,c)=\frac 12 k_{\mbox{\tiny{B}}}T \bigg ( v
\rho^2+ \int\frac{d^d{\bf k}}{(2\pi )^d} \log \Big (1+\frac
vV S_0({\bf k},c) \Big)\bigg)~,
\end{equation}
where $\rho=N/V$ is the mean physical density. Note that no
further averaging over the frozen crosslink positions is
needed, since $S_0({\bf k},c)$ in (23) was already shown to
be self-averaging.  For further reference we quote also the
result of the mean field free energy in terms of mean
density fluctuations $\langle\Phi_{\bf k}\rangle$.  Up to
an irrelevant constant it has been shown \cite{doied} that
\begin{equation} 
F(v,c)=\frac 12 k_{\mbox{\tiny{B}}}T \int\frac{d^d{\bf
k}}{(2\pi )^d}\, S^{-1}({\bf k},c) \langle\Phi_{\bf
k}\rangle \langle\Phi_{-{\bf k}}\rangle ~,
\end{equation}
where $S^{-1}({\bf k},c)=(v/V)+S_0^{-1}({\bf k},c)$ is the
inverse structure function of the interacting system in
Gaussian approximation.  Higher order terms in the above
Landau type expansion are readily calculated from equation
(20) by modifying ${\bf c}$ in (22) accordingly, but will
be neglected in the following discussion.

\subsection{Crosslinked polymer blends}

Interpenetrating polymer networks (IPNs), semi-IPNs and
crosslinked polymer blends have  broad range of
applications in polymer research and material sciences
\cite{vilcol,sperli}. There has been some effort to model
these systems using a propagator originally proposed by de
Gennes \cite{degen} which led to physically reasonable
results and agreed widely with experiment \cite{ipnvil}.
However, these were semi-phenomenological models and a
microscopic theory for multi-component networks is to the
best of our knowledge still lacking.

We consider the case of a highly crosslinked two-component
polymer blend.  For sufficiently strong incompatibility
between the network components we expect phase separation
on a mesoscopic length scale  in which A rich and B rich
microdomains are formed. To model such a situation we start
from a long A-B diblock copolymer with the following
structure
\begin{equation}
{\bf R}_i=\left\{
\begin{array}{ll}
\mbox{A monomer} & \mbox{if } ~~~0 \leq i < \phi N ~, \\
\mbox{B monomer} & \mbox{if } ~\phi N \leq i \leq N~. 
\end{array}
\right . 
\end{equation}
Here $\phi$ and $1-\phi$ are the volume fractions of the A
and B components. Crosslinks between monomers are
introduced in the same fashion as in equation (4). The
additional crosslink between the A and B chain has of
course no severe consequences. The interaction is described
by a proper generalization of $H_{\mbox{\tiny I}}$ in (2)
\begin{eqnarray}
\beta H_{\mbox{\tiny I}} &=&
\frac{1}{2} \Big (
v_{\mbox{\tiny AA}}\sum_{i,j=0}^{\phi N-1} \delta ({\bf
R}_i-{\bf R}_j)+ v_{\mbox{\tiny BB}}\sum_{i,j=\phi N}^{N}
\delta ({\bf R}_i-{\bf R}_j)\nonumber\\ &+& 2
\,v_{\mbox{\tiny AB}}\sum_{i=0}^{\phi N-1}\sum_{j=\phi N}^N
\delta ({\bf R}_i-{\bf R}_j) \Big)~.
\end{eqnarray}
We follow here closely Leibler's derivation \cite{leibler}
for diblock copolymer melts. Using the more general
expression (29) in the Hamiltonian (5), the free energy is
calculated in terms of one component density fluctuations
$\Phi_{{\bf k}}^{\mbox{\tiny A}}=\sum_{i=0}^{\phi N-1}\exp(
i{\bf k}{\bf R}_i)$ up to second order.  From this standard
calculation \cite{benoit}  a free energy functional of the
same form as (27) is obtained with a modified expression
for the structure function $S$ \cite{leibler}. For an
incompressible system of symmetric copolymers
($v_{\mbox{\tiny AA}}=v_{\mbox{\tiny BB}}=v$ and
$v_{\mbox{\tiny AB}}=v+\Delta v$) $S$ is given in terms of
crosscorrelation functions of single component density
fluctuations $S_0^{xy}({\bf k},c)=\langle
\Phi_{\bf k}^x\Phi_{-\bf k}^y \rangle_0$ ($x,y=$ A,B) and
the Flory parameter $\chi_{\mbox{\tiny F}}=\rho\Delta v$
\cite{leibler}
\begin{equation}
S^{-1}=\frac{S_0^{\mbox{\tiny AA}}+S_0^{\mbox{\tiny
BB}}+2S_0^{\mbox{\tiny AB}}}{S_0^{\mbox{\tiny
AA}}S_0^{\mbox{\tiny BB}}-(S_0^{\mbox{\tiny AB}})^2} -2
\frac{\chi_{\mbox{\tiny F}}}{N}~. 
\end{equation}
With the method described in section 4 it is
straightforward to calculate $S_0^{xy}({\bf k},c)$ by a
proper generalization of (23) for arbitrary volume fraction
$0\leq \phi\leq 1$.  In the disordered phase it is
sufficient to consider quadratic terms of the order
parameter $\langle \Phi_{{\bf k}}^{\mbox{\tiny A}}\rangle $
in the Landau free energy, equation (27). A complete
understanding of microphase separation would require the
study of the homogeneous phase as well as the ``ordered''
mesophase. As precursor for phase instability we analyze
the behaviour of the structure function $S$, i.e., the
correlation function of the local order parameter, at its
maximum. From (30) it is clear that the position of the
maximum can only depend on the magnitude of ${\bf k}$ and
is independent of $\chi_{\mbox{\tiny F}}$. From our
calculation we find a maximum of $S$ at finite wavevector
${\bf k}_0$ (figure 3). Divergency of the structure
function in (30) (i.e., $S^{-1} \rightarrow 0$) at its peak
value ${\bf k}_0$ indicates instability of the disordered
phase and serves as definition for critical
$\chi_{\mbox{\tiny c}}$ (figure 3).  More calculational
details will be presented elsewhere.  For a lucid
discussion of the above method, see reference
\cite{leibler}.

\section{Outlook and conclusion}

There are many directions to extend this work and open
problems in connection with macromolecular networks. Here
we  mention some of them mainly for completeness.

In this paper we have considered networks that were
modelled from a single chain of macroscopic size.  While
this simplification is physically reasonable for highly
crosslinked polymers \cite{deamed,edwa,badoed}, it
completely fails to describe the correct physics of dilute
systems. The problem of weakly crosslinked polymers is of
special relevance in connection with the vulcanization
transition, i.e., the transition from a liquid to an
amorphous solid state. Upon increasing crosslink
concentration the gel-fraction of the network grows until
at the percolation threshold an infinitely large network
cluster is formed. In this case it was shown that a finite
fraction of polymers spontaneously localizes \cite{golzip}
with respect to some reference frame (e.g., the center of
mass of the gel) and therefore can support applied stress.
Although we are not dealing here with the vulcanization
problem ({\it all} monomers are {\it a-priori} localized
via the harmonic potentials, equation (4)), it is
instructive to see how the same issue arises in our
formulation.

In a dilute network consisting of many unconnected or
partially connected polymer chains, each network cluster
has its own center of mass mode.  In this case it is no
longer possible to integrate over each center of mass
coordinate separately as was done in the beginning of
section 3.  The simplifying feature there was of course
simply-connectedness of the object.  Since any further
progress required the matrix in (14) to be positive
definite, a randomly crosslinked dilute network consisting
of many chains poses additional mathematical problems.  It
is easy to see that for multiple chains, equation (A1) has
to be replaced by the more general expression for a polymer
melt
\begin{equation}
{\cal I}_n \otimes \left(
\begin{array}{cccc}
0 & 0 & \cdots & 0 \\ 0 & 1 & \cdots & 0 \\
\vdots & \vdots & \ddots & \vdots \\ 
0 & 0 & \cdots & 1
\end{array}
\right )~. 
\end{equation}
Here ${\cal I}_n$ is the $n$ dimensional unit matrix, $n$
the number of polymer chains, and $\otimes$ denotes a
direct product. Equation (31) has a simple physical
interpretation. The 1's represent the backbone of the chain
and by removing one element from the diagonal, the chain is
cut into two pieces, and so forth. It is interesting to
note that even for this more general situation, it is still
possible to ``invert'' ${\cal M}$ analytically, although it
is no longer positive definite.  This was proven in
reference \cite{lewnew} by making use of the concept of
pseudoinverse matrices which leads to a similar result as
equation (18).  However, the mathematics is more involved
and will not be presented in this paper.  In the framework
of replica field theory the vulcanization problem was
treated in references \cite{golzip} by invoking a
variational ansatz for the localization length of monomers.

Finally we want to make a short comment about the role of
entanglements in polymeric networks.  Up to now the systems
considered were phantom in a sense that the only
topological restrictions on chain motion were permanent
crosslinking junctions modelled by (4). In any realistic
vulcanization process  upon network formation a certain
number of entanglements is permanently trapped which can
be viewed as another form of frozen constraint leading to
reduction of entropy.  However, up to now there are no
topological invariants known to mathematicians that
describe this phenomenon in an unique and rigorous fashion.
A simplified version of the entanglement problem was
proposed by Edwards and coworkers who have modelled
entanglements in form of sliplinks \cite{badoed,edvil},
i.e., crosslinks with the additional freedom to slide along
the chain.  Formally sliplinks are introduced by treating
the crosslink coordinates as ``hot'' variables. To see how
equation (3) gets modified in the presence of sliplinks, we
consider the simplest possible scenario in which all
crosslinks in (3) are assumed to be sliplinks.  To model a
more realistic situation we had to consider both,
crosslinks and sliplinks. Summing (3) over all $i_e,j_e$
independently modifies the former crosslink term to $\big
(\sum_{i,j=0}^N\delta \left( {\bf R}_i-{\bf R}_j\right)\big
)^M$. By invoking the well known identity $x^M=(M!/2\pi
i)\oint d\mu \,e^{\mu x}\, \mu^{-(M+1)}$, it becomes clear
that the sliplink contribution can be treated on a similar
footing as the excluded volume interaction in (2). A
mathematically similar but more involved problem arises in
the replica formalism, and it had been shown \cite{deamed}
that the $\mu$-integration is dominated by the steepest
descent. Thus in the simple example considered the
effective excluded volume is reduced by the presence of
sliplinks to $v-\mu$. The above discussion is also valid
for independently distributed crosslinks and sliplinks.
Whereas the crosslinks modify the Wiener measure to (4),
sliplinks renormalize the excluded volume parameter. More
sophisticated models can be formulated if the degrees of
freedom of the sliplinks are in addition restricted by the
permanent crosslink positions \cite{badoed,edvil}.\\

In summary, we have proposed a microscopic model for RCGNs
and its generalization to interacting networks and
crosslinked polymer blends. As an extension of the Edwards
Hamiltonian for linear polymers we introduced a similar
Hamiltonian for RCGNs taking excluded volume interaction
into account.  We developed a mathematical formalism which
can solve the non-interacting RCGN exactly. By employing
the idea of self-averaging we showed how to approach the
excluded volume problem in a systematic, perturbative
manner.  Our method avoids the well known technical
difficulties of replica theory like replica symmetry
breaking or the $n\rightarrow 0$ limit. The presented
theory provides a new perspective on various aspects and
open questions of polymer networks with random and
non-random connectivities that are of general interest.

\acknowledgments 
Financial support of this work by the Deutsche
Forschungsgemeinschaft, SFB 262, is gratefully
acknowledged.

\appendix 

\section{}

To prove equation (14) we have to evaluate ${\cal D}^t
\big (z{\cal W}+\sum_{e=1}^M {\cal K}(i_e,j_e) \big ){\cal
D}$ in (9) with matrices defined via (10--12). First note
that
\begin{equation}
{\cal D}^t{\cal W}{\cal D}=
\left(
\begin{array}{cccc}
0 & 0 & \cdots & 0 \\ 0 & 1 & \cdots & 0 \\
\vdots & \vdots & \ddots & \vdots \\ 
0 & 0 & \cdots & 1
\end{array}
\right )~. 
\end{equation}
Let ${\bf e}_i=(0,...,0,1,0,...,0)^t$ be the $i$th
unitvector of the canonical basis in $N+1$ dimensional
space. Then
\FL
\begin{eqnarray}
\sum_{e=1}^M{\cal K}(i_e,j_e)&=&\sum_{e=1}^M \Big({\bf
e}_{i_e}^{}{\bf e}_{i_e}^t+{\bf e}_{j_e}^{}{\bf
e}_{j_e}^t-{\bf e}_{i_e}^{}{\bf e}_{j_e}^t-{\bf
e}_{j_e}^{}{\bf e}_{i_e}^t \Big )\nonumber \\ &=& 
\sum_{e=1}^M \big ({\bf e}_{i_e}^{}-{\bf e}_{j_e}^{}\big)
\big( {\bf e}_{i_e}^{}-{\bf e}_{j_e}^{}\big)^t \nonumber \\
&=& \Big( ({\bf e}_{i_1}^{}-{\bf e}_{j_1}^{}),..., ({\bf
e}_{i_M}^{}-{\bf e}_{j_M}^{})\Big )
\Big( ({\bf e}_{i_1}^{}-{\bf e}_{j_1}^{}),...,
({\bf e}_{i_M}^{}-{\bf e}_{j_M}^{})\Big )^t~.
\end{eqnarray}
From (A2)  we obtain equations (14) and (15) by setting
${\bf p}_e=-{\cal D}^t({\bf e}_{i_e}^{}-{\bf e}_{j_e}^{})$.

\section{} 

We want to find the inverse of the matrix defined in
equation (14),
\begin{equation}
{\cal M}(z)=z\Big ({\cal I}+\frac 1z {\cal P}{\cal P}^t\Big
)~, 
\end{equation} 
in the limit $z \rightarrow 0$. The difficulty with (B3) is
that ${\cal I}$ cannot be neglected against the $1/z$ term,
since the crosslink contribution alone is highly singular
with  degeneracy $N-M$.  We proceed by writing the inverse
of ${\cal M}$ quite formally in terms of its Neumann series
\begin{eqnarray}
{\cal M}^{-1}(z)&=&\frac 1z \Big ({\cal I}+\frac 1z {\cal
P}{\cal P}^t\Big )^{-1}=\frac 1z \sum_{n=0}^\infty \big
(-\frac 1z {\cal P}{\cal P}^t\big)^n\nonumber \\ &=&\frac
1z \bigg ({\cal I}-\frac 1z {\cal P} 
\Big ( \sum_{n=0}^\infty \big (-\frac 1z {\cal P}^t{\cal
P}\big)^n \Big ){\cal P}^t\bigg )\nonumber \\
&=&\frac 1z \bigg ({\cal I}-\frac 1z {\cal P} \Big ({\cal
I}+\frac 1z {\cal P}^t{\cal P}\Big)^{-1}{\cal P}^t\bigg )~.
\end{eqnarray}
With the definition in (15) it is easy to show that for
tetrafunctional crosslinking junctions, the crosslink
matrix ${\cal P}$ is of full rank $M$.  It is well known
that in this case the normal form ${\cal P}^t{\cal P}$ is
positive definite \cite{zurfa} and that therefore its
inverse exists.  Using this information we obtain equation
(18) from (B4) by letting $z\rightarrow 0$. Convergency of
the above manipulations is proved by multiplying the final
result for ${\cal M}^{-1}$ with ${\cal M}$ in (B3) which
gives identity.  This also verifies the correctness of the
result in (B4).

\begin{figure}
\caption{A single-chain network. Only the beads at the
crosslink sites are shown. For high crosslink
concentrations far above the percolation threshold a
single-chain network is practically indistinguishable from
a network that is modelled from many polymer chains.}
\end{figure}

\begin{figure}
\caption{Scattering function $S_0(q,C)$ for two different
RCGNs derived from (23). $q^2={\bf k}^2a^2/2d$; $c=M/N$;
solid lines $N=10000$, $c=0,\,1,\,2,\,3,\,4,\,5\, \%$
(left to right); open circles $N=5000$ (
only shown for $c=2,\,4\, \%$). 
$\mbox{C}=\{i_e,j_e\}_{e=1}^M$ were picked from the
interval $[0,N]$ at random. The orthonormalization of
${\cal P}$ was performed with a standard singular value
decomposition algorithm [16].  Due to self-averaging curves
with same crosslink concentration (here $c=2,\,4\,\%$) are 
identical.  For $c\rightarrow 0$ we recover the 
Debye function for linear polymers.}
\end{figure}

\begin{figure}
\caption{Structure function (30) of a symmetric crosslinked
polymer blend for three different values of the Flory
parameter $\chi_F=0,\,2c,\,4c$ (upwards) with $c=M/N=0.02$
($M=200$, $N=10000$) and $\phi =0.5$. We find that a
maximum of $S$ occurs at finite wavevector $q_0=0.23$
($q^2={\bf k}^2a^2/2d$). For the above values of $\phi$ and
$c$ the disordered phase becomes unstable for $\chi_F \geq
\chi_c \approx 4.3c$.}
\end{figure}
\end{document}